\documentstyle[epsfig]{aipproc}
\begin{document}
\title{Calibration of the CAT Telescope}

\author{Fr\'ed\'eric Piron, for the CAT collaboration}
\address{Laboratoire de Physique Nucl\'eaire des Hautes Energies\\
Ecole Polytechnique, route de Saclay, 91128 Palaiseau Cedex, France}

\maketitle

\begin{abstract}
Due to the lack of test-beams in ground-based $\gamma$-ray astronomy, detector calibration has been a major
challenge in this field. However, with the use of Cherenkov ring-images due to cosmic-ray muons and of strong
$\gamma$-ray signals, the CAT telescope could be
rather well monitored and understood. Here we present a few oustanding aspects of this work.
\end{abstract}
\section*{Introduction: the CAT detector}
\begin{figure}[t!]
\centerline{
\hbox{
\hspace{-.5cm}
\epsfig{file=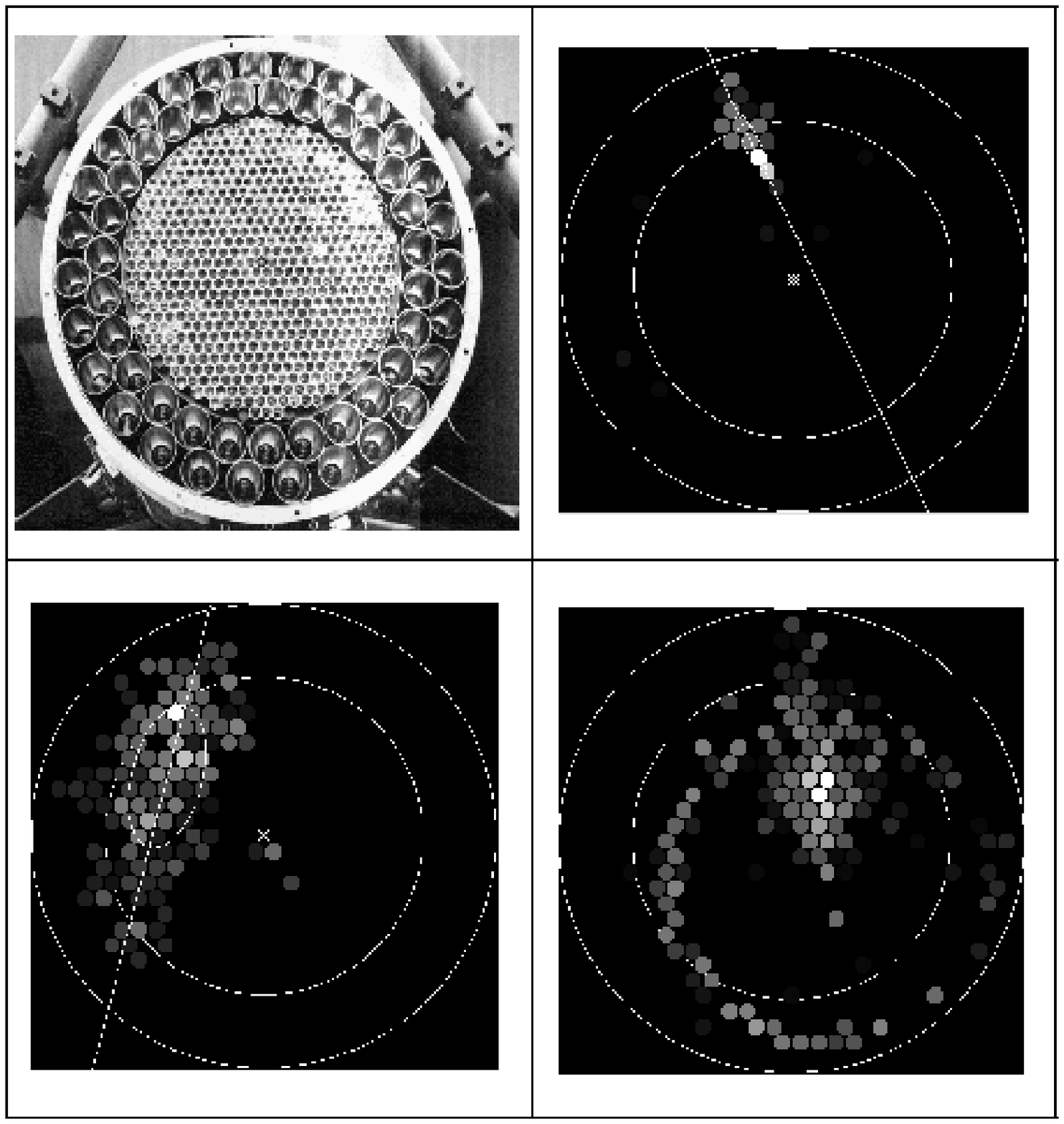,width=6cm,height=6cm,clip=
,bbllx=80pt,bblly=245pt,bburx=535pt,bbury=730pt}
\epsfig{file=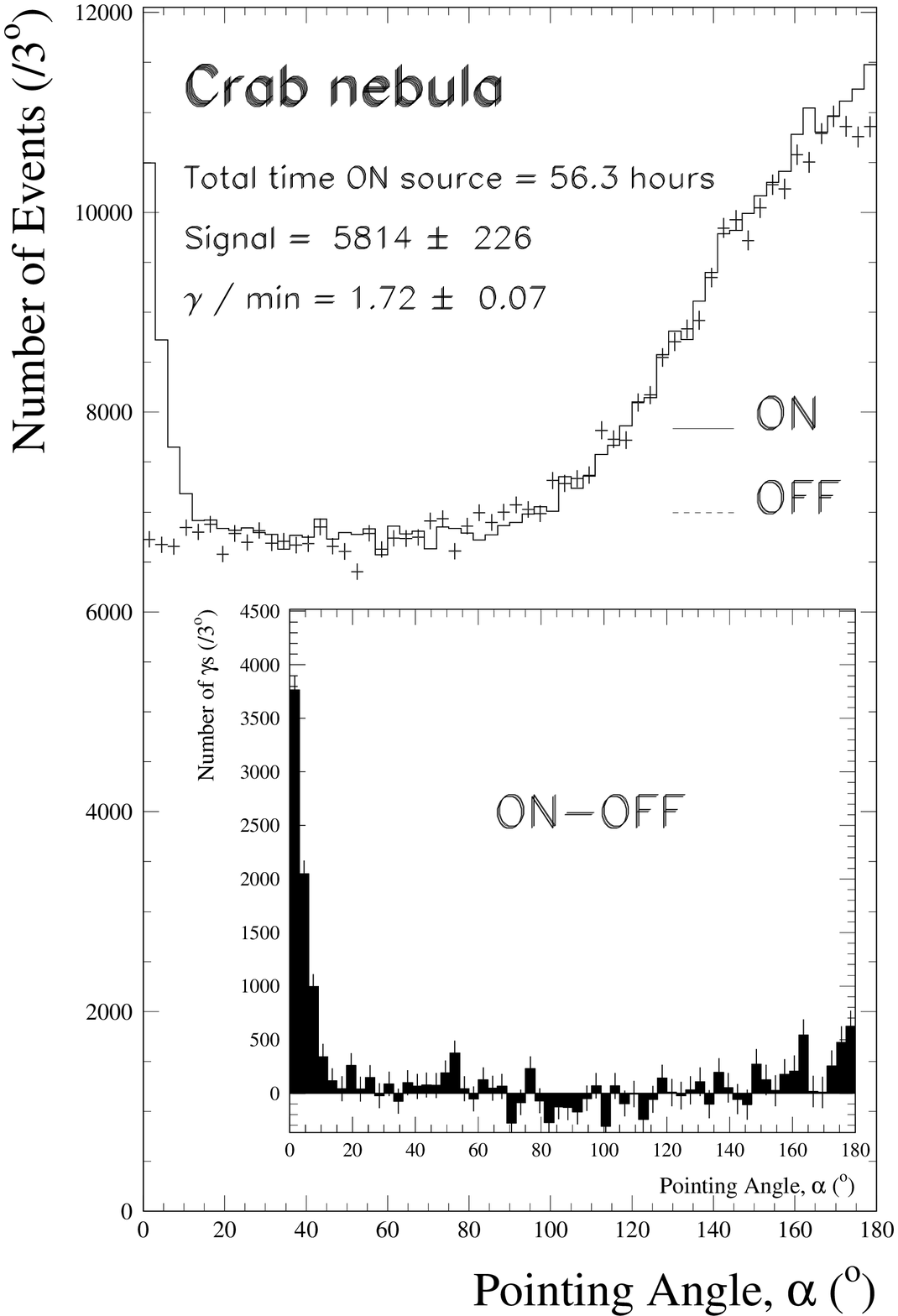,width=8cm,height=6cm,clip=
,bbllx=5pt,bblly=5pt,bburx=500pt,bbury=715pt}
}}
\caption{
{\it (a)} The CAT camera and three typical events:
the first image is presumably due to a $\gamma$ shower because of its fine and cometary shape, and because it is
pointing towards the source position at the center of the camera; the second image is
certainly due to a hadron because it is more spread out and pointing elsewhere; finally, the third image is clearly
hadronic and would be easily rejected, because it is signed by the presence of a muon ring.
{\it (b)} Pointing angle distribution from a sample of data taken on the Crab nebula between 1996 and
1999, for zenith angles ranging from 21$^\circ$ to 35$^\circ$. Cuts on total charge and shape have been applied (see text).
The inset shows the ON$-$OFF distribution (OFF source runs are used to estimate the hadronic background):
%
%
within 6$^\circ$, the total significance is 25.6$\sigma$ for a ratio of durations
${\mathrm{T_{ON}}}/{\mathrm{T_{OFF}}}=2.4$ (the OFF distribution has been renormalized). The corresponding
significance for an equal amount of ON and OFF runs would be 32.2$\sigma$, i.e. 4.3$\sigma$ in one hour.}
\label{sample}
\end{figure}

The C{\small AT} (Cherenkov Array at Th{\'e}mis) telescope records Cherenkov flashes due to VHE atmospheric
showers through its 17.8m$^2$ mirror. Its camera is located 6m from the mirror and has a 4.8$^\circ$ full field of
view, consisting of a central region of 546 0.12$^{\circ}$ angular diameter phototubes arranged in a
hexagonal matrix and of 54 surrounding tubes in two ``guard rings'' (Fig.\ref{sample}a).
Fast electronics allows a relatively low $\gamma$-ray detection threshold energy
of $250\:\mathrm{GeV}$ (at Zenith), and the fine grain of the camera permits an accurate image analysis.
The experiment and the analysis method are fully described elsewhere~\cite{Barrau,LeBohec}.
Briefly, after selecting the most significant triggers (total charge $Q_{\mathrm{tot}}$$>$30 photo-electrons), good
discrimination between $\gamma$ and hadron-induced showers is achieved by looking at the shape and the orientation
of the images (see the events on Fig.~\ref{sample}a): since $\gamma$-ray images are rather thin and ellipsoidal
while hadronic images are more irregular, a first cut is applied which selects images with a ``$\gamma$-like''
shape; it is based on a $\chi^2$ fit to a mean light distribution predicted from electromagnetic showers, and a
probability $P({\chi^2})$$>$0.35 is required. Then, since $\gamma$-ray images are expected to point towards the
source position
in the focal plane whereas cosmic-ray directions are isotropic, a second cut $\alpha$$<$6$^{\circ}$ is used in the
case of a point-like source, where the pointing angle $\alpha$ is defined as the angle at the image barycentre
between the actual source position in the focal plane and that of the image which is reconstructed by the
fit~\footnote{The resolution per event is of the order of the pixel size, i.e. $\sim$0.1$^\circ$.}.
As a result, this procedure rejects 99.5\% of hadronic events while selecting 40\% of
$\gamma$-ray events. Fig~\ref{sample}b is an example of the pointing angle distribution obtained on the
Crab nebula: the signal is clearly seen in the first bins, while a second signal can be seen at
$\alpha$$\sim$180$^\circ$,
due to $\gamma$-ray images whose direction has been mis-reconstructed by the fit~\footnote{The global rise of
the background distribution for large values of $\alpha$ corresponds to large hadronic images which
were cut by the edge of the camera. This effect becomes fainter for larger zenith angles, since images form
closer to the center of the camera; see~\cite{Mohanty} for illustration.}.
\section*{Calibration of the detector} 
\subsection*{Hardware monitoring}
\begin{figure}[t!]
\centerline{
\hbox{
\epsfig{file=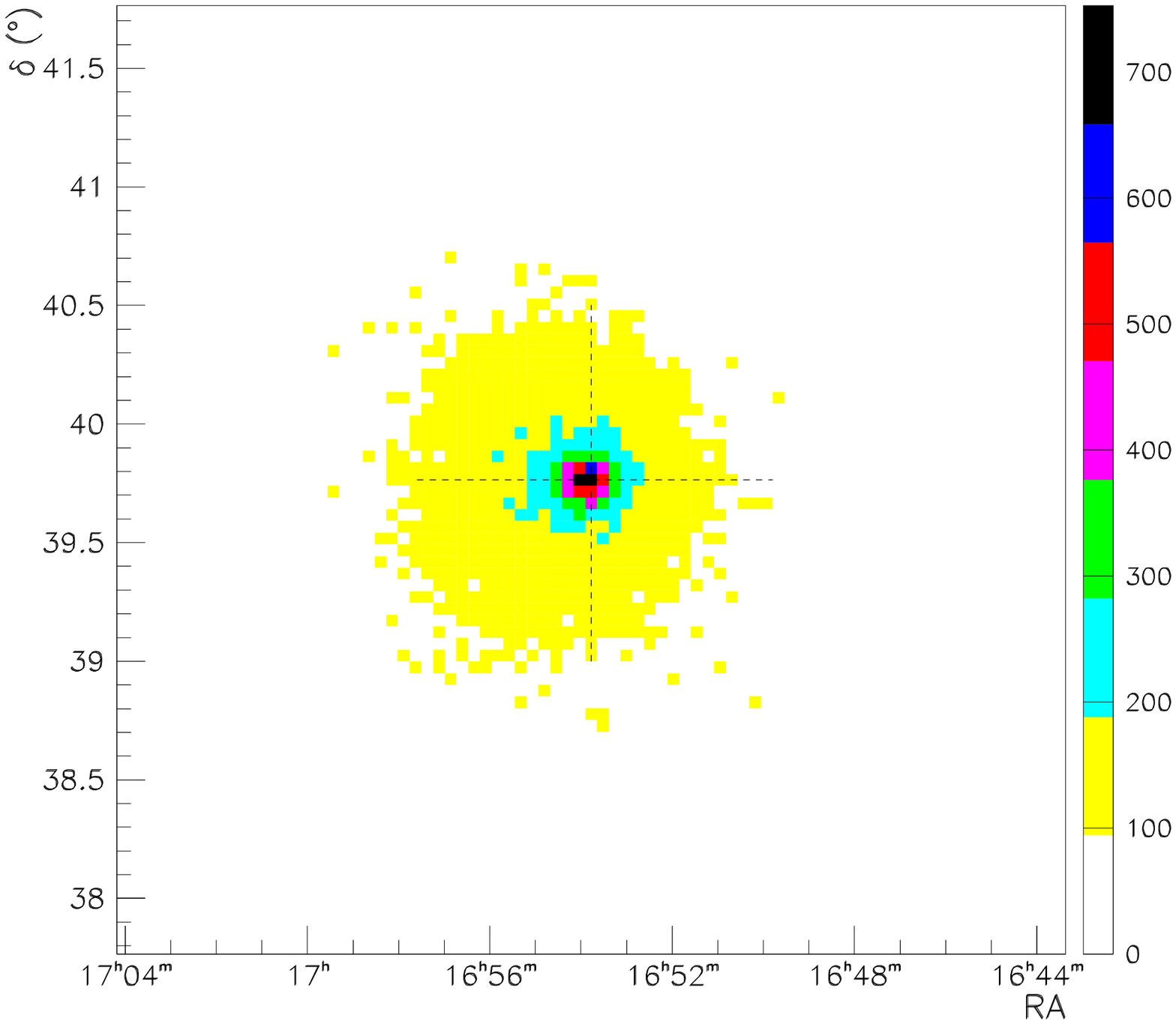,width=4cm,height=3.5cm,clip=
,bbllx=1pt,bblly=20pt,bburx=563pt,bbury=515pt}
\epsfig{file=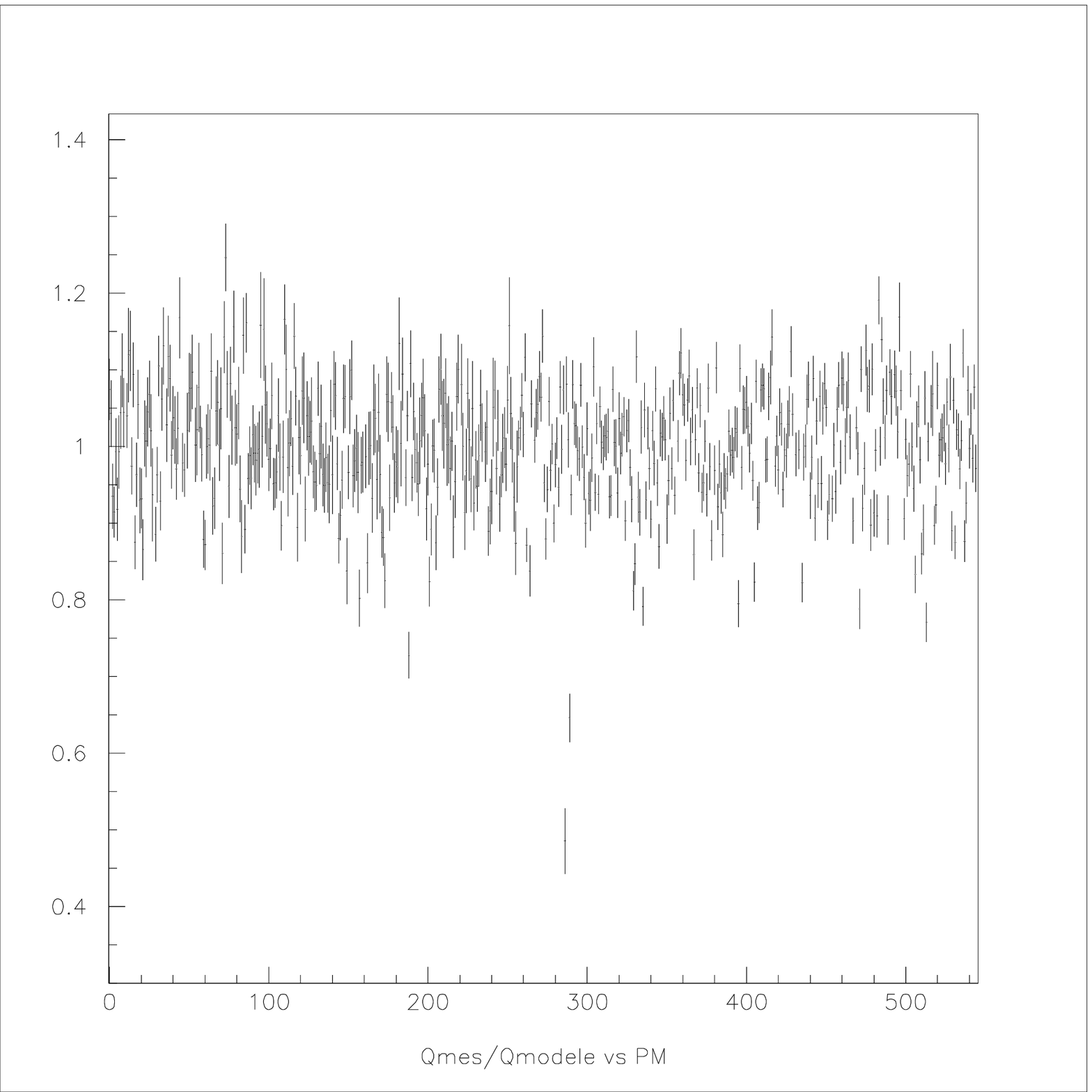,width=10cm,height=3.5cm,clip=
,bbllx=18pt,bblly=3pt,bburx=516pt,bbury=518pt}
}}
\caption{
{\it (a)} Projected distribution (bin size=0.05$^\circ$) of the reconstructed shower angular origins for all data
of Mrk~501 in 1997: no background substraction has been performed, and events are selected by the shape cut only.
The cross marks the actual position of Mrk~501.
{\it (b)} Ratio of the charge integrated in each of the 546 small inner phototubes to that predicted by a model
of muon rings. $\sim$100 muon images have been used for these statistics. The dispersion of the ratio around
unity is mainly due to the uncertainty on the gain values.}
\label{mechopt}
\end{figure}
Mrk~501 exhibited a remarkable series of flares during the whole year 1997~\cite{Djannati}, which have been very
useful for detector calibration. As an example, Fig.~\ref{mechopt}a illustrates the very good quality of the
mechanical monitoring for data taken with a zenith angle between 0$^\circ$ and 44$^\circ$: the signal appears
right at the actual position of the source, thus validating the angular correction which is applied on each event
to compensate for the unavoidable and zenith-dependent slight mis-alignment of the optical axis of the 
camera~\footnote{This is due to the weight of the camera, which bends the arms of the telescope, and to the telescope
azimuth and altitude axis mis-alignment, especially at large zenith angles.}.\\
\indent Cosmic-ray muons falling onto the mirror yield ring-like images in which the light distribution can be
easily predicted. The fine grain of the C{\small AT} camera allows a fine analysis of these
images~\cite{IacoucciThese}.
In this way, the overall conversion factor between ADC counts and incident Cherenkov photon number can be directly
checked. This factor involves optical efficiencies, as well as phototube pedestals and gains. As an example,
Fig~\ref{mechopt}b shows that the camera is correctly calibrated, except for a few channels which are not used in
the analysis. In this study, particular attention has been paid to the wavelength-dependent aspect, by
taking special runs using different UV-filters placed in front of the camera.
\subsection*{Validation of the simulations}
\begin{figure}[t!]
\centerline{
\hbox{
\epsfig{file=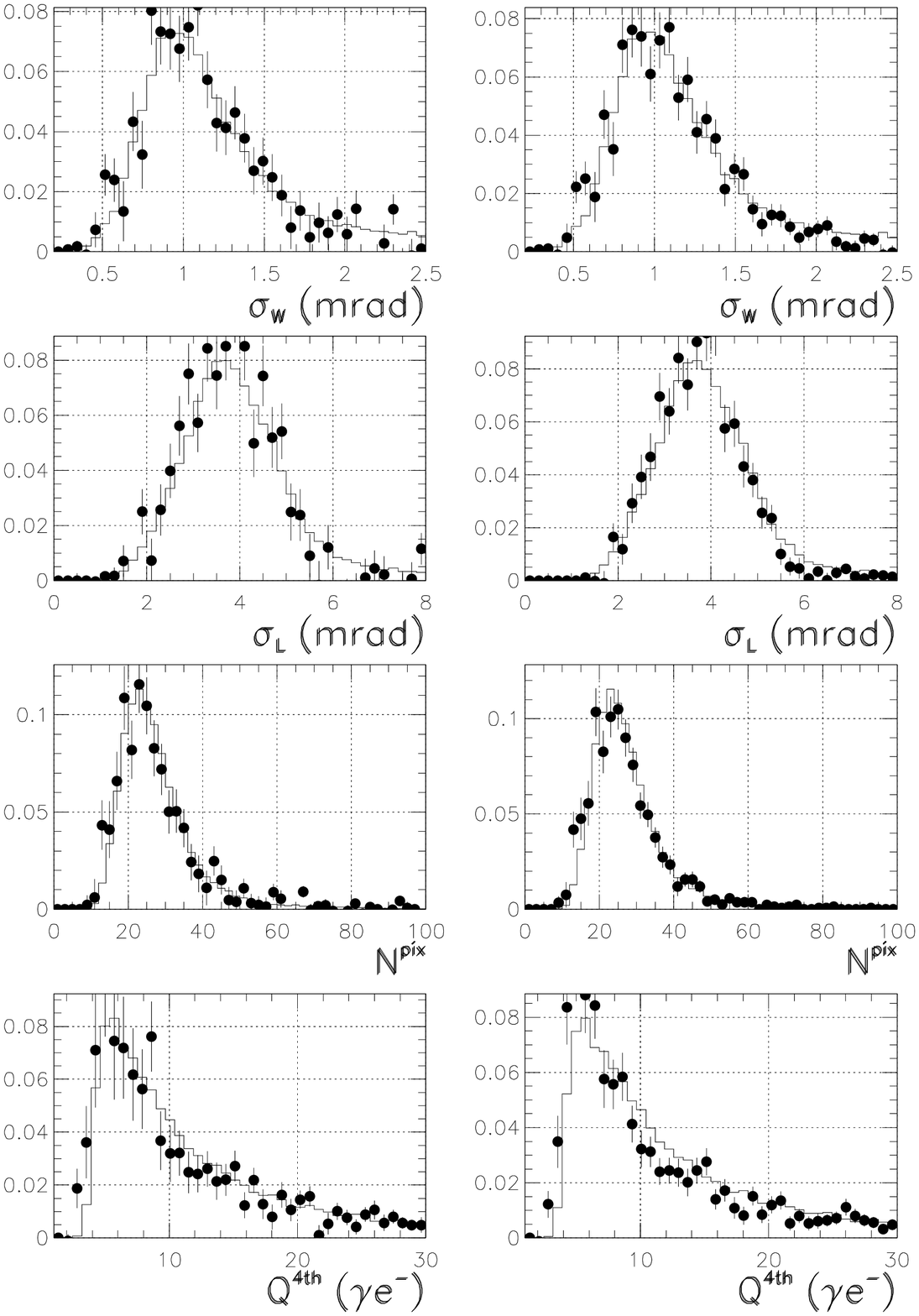,width=7cm,height=6.5cm,clip=
,bbllx=20pt,bblly=15pt,bburx=550pt,bbury=765pt}
\epsfig{file=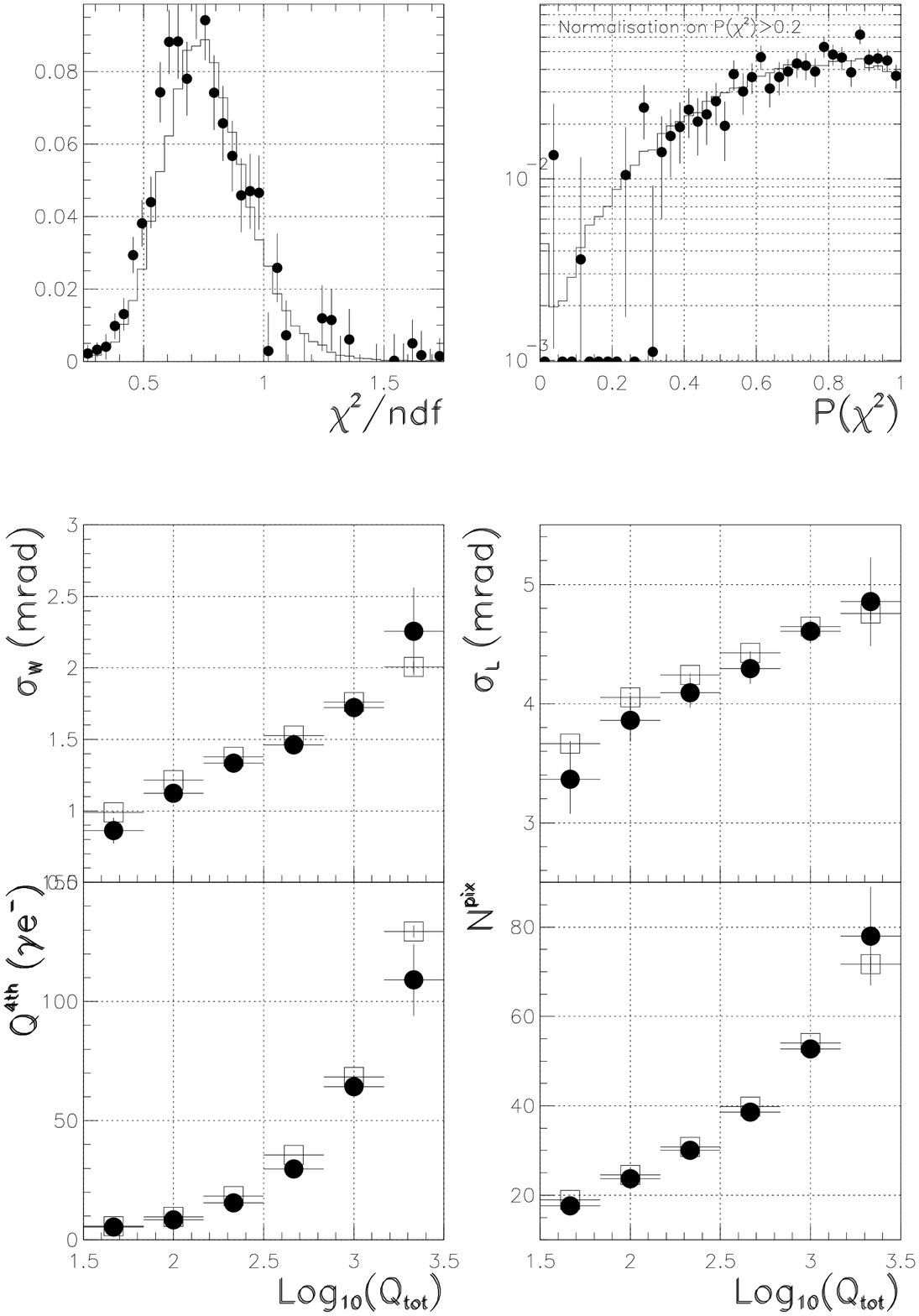,width=7cm,height=6.5cm,clip=
,bbllx=5pt,bblly=10pt,bburx=550pt,bbury=790pt}
}}
\caption{Comparison of $\gamma$-ray shower simulations (full lines or open squares) with the data (ON$-$OFF) from
the flare of Mrk~501 on April 16$^{\mathrm{th}}$, 1997 (filled points). Statistical errors are negligible for
simulations.
{\it (a)} Distribution of several variables, within the orientation cut $\alpha$$<$6$^{\circ}$ (left
column) and with the cut on shape $P({\chi^2})$$>$0.35 also (right column):
Hillas parameters {\it width} ($\sigma_{W}$) and {\it length} ($\sigma_L$), number of pixels in the image
$N^{\mathrm{pix}}$, and fourth-brightest-pixel's charge $Q^{\mathrm{4th}}$;
{\it (b)} $\chi^2/$ndf and $P({\chi^2})$ distributions, and previous parameters as functions of
Log$_{10}(Q_{\mathrm{tot}})$.}
\label{mcdata}
\end{figure}
\begin{figure}[b!]
\centerline{
\hbox{
\epsfig{file=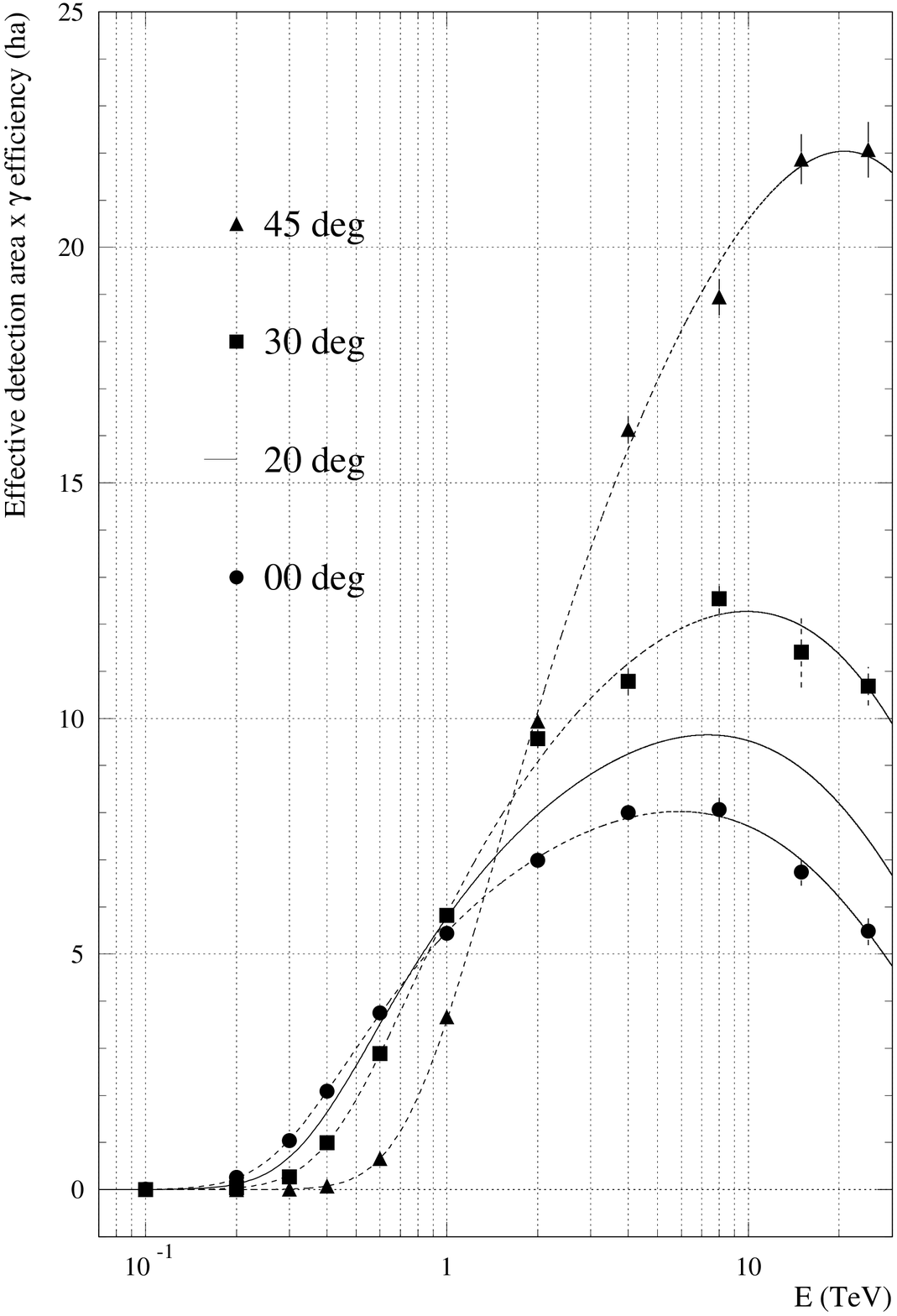,width=5cm,height=5.5cm,clip=
,bbllx=25pt,bblly=40pt,bburx=540pt,bbury=795pt}
\epsfig{file=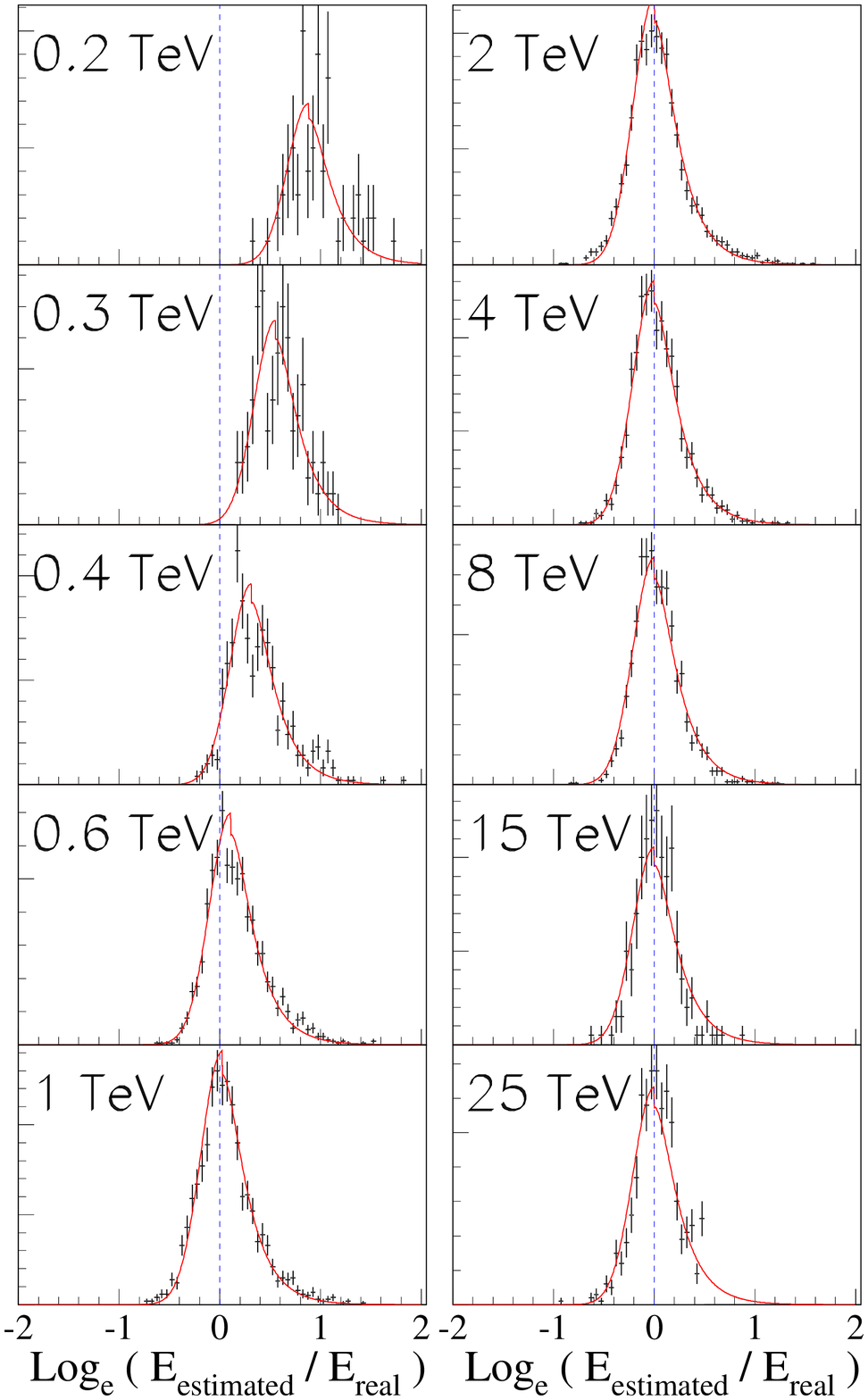,width=9cm,height=5.5cm,clip=
,bbllx=74pt,bblly=32pt,bburx=540pt,bbury=790pt}
}
}
\caption{
{\it (a)} $\gamma$-ray effective detection area (in 10$^4$m$^2$), including the effect of event-selection
efficiency, as a function of the energy. Each point corresponds to simulations, while full lines come from an
analytical 2D-interpolation over energy and zenith angle;
{\it (b)} Energy-resolution functions $\Psi(E\rightarrow\widetilde{E},cos\theta)$ {\it vs} $\log(\widetilde{E}/E)$, for
a fixed zenith angle $\theta=30^\circ$ ($y$-axis units are arbitrary). Each plot is defined for a fixed value of
the {\it real} energy $E$, ranging from $200\:\mathrm{GeV}$ to $25\:\mathrm{TeV}$, and runs on the {\it estimated}
energy $\widetilde{E}$. Full lines come from an analytical 3D-interpolation over the three variables $E$,
$\widetilde{E}$ and $\theta$.}
\label{accres}
\end{figure}
On April 16$^{\mathrm{th}}$, 1997, Mrk~501 reached $\sim$8 times the level of the Crab nebula.
The signal-to-noise ratio for this night is 2.7, corresponding to a $\gamma$-ray beam with only 30\% contamination.
The good atmospheric quality of this night allows the use of this beam for calibration through comparison
with simulations. Fig.~\ref{mcdata}a shows the perfect agreement on the distributions of Hillas
parameters~\cite{Weekes}: it is shown for events selected both with the single orientation cut and with the
complete selection including the $\chi^2$ fit. The distribution of the final number of pixels $N^{\mathrm{pix}}$
retained by the fit (see~\cite{LeBohec}) is also well reproduced. Furthermore, the good agreement
observed on the fourth-brightest-pixel's charge $Q^{\mathrm{4th}}$ validates the simulations at the trigger level,
since the trigger condition requires four pixels above threshold.
Finally, Fig.~\ref{mcdata}b shows the $\chi^2/$ndf and $P({\chi^2})$
distributions~\footnote{The $P({\chi^2})$ distribution
is not flat, as would be expected if the $\chi^2$ were performed using variables with Gaussian
errors. The description of Cherenkov light fluctuations in showers development is a very difficult
task (see~\cite{LeBohec} for discussion).}, as well as the previously discussed parameters, expressed as functions
of the total image charge $Q_{\mathrm{tot}}$: here again the agreement is very good. This allows the simulation
to be used to calculate the $\gamma$-ray effective
detection area within the selection cuts, as well as the energy-resolution function $\Psi$. This is shown in
Fig.~\ref{accres}. In particular, a clear positive bias in the energy reconstruction is visible for low values of
the injected energy (Fig.~\ref{accres}b): it is due to a trigger effect which selects those events which benefited
from a positive fluctuation of Cherenkov light during the shower development. This energy over-estimation disappears
when going towards higher energies, where $\Psi$ becomes more Gaussian, with a zero mean value and a width
$\sigma$$\sim$20\%.
\begin{figure}[t!]
\centerline{
\hbox{
\epsfig{file=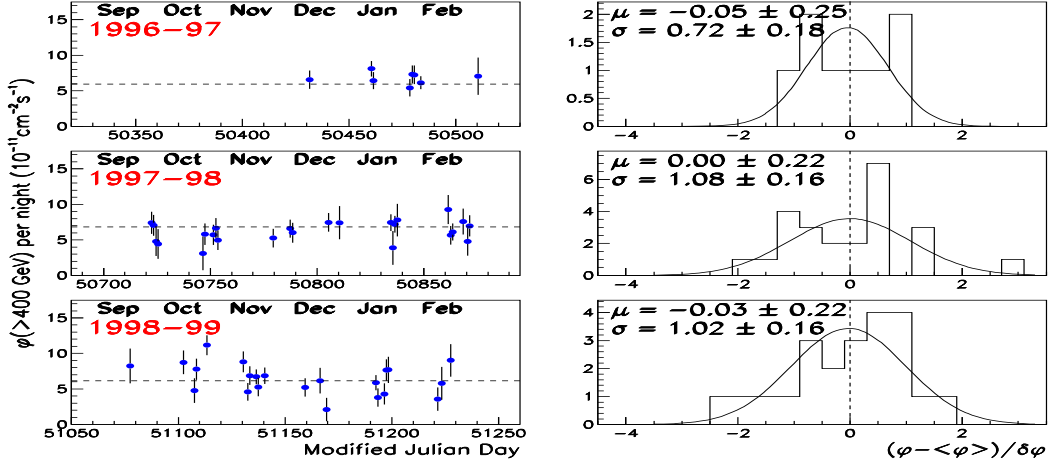,width=14cm,height=6.2cm,clip=
,bbllx=0pt,bblly=5pt,bburx=530pt,bbury=375pt}
}}
\caption{1996 to 1999 nightly integral flux of the Crab nebula above $400\:\mathrm{GeV}$, from a sample of data with
zenith angles up to 35$^\circ$. The dotted lines represents the mean flux for each period, and the corresponding
residuals are shown on the right panel.}
\label{crabstab}
\end{figure}
\section*{Conclusion}
More details concerning calibration and spectrum measurement will be given in a forthcoming paper. The good
stability obtained to date on the signal from the Crab nebula (Fig.~\ref{crabstab}), using the acceptances and
energy-resolution function discussed above, illustrates the good quality of running conditions and the good
understanding of the detector.
Future effort will be devoted to the analysis of large zenith angle data (see~\cite{Mohanty} for a first study) and
to the cross-calibration with the C{\small ELESTE} experiment, operating on the same site
with an energy threshold of $\sim$$50\:\mathrm{GeV}$; the recent observation of the first common events between both
detectors~\cite{DeNaurois} is an encouraging result in this direction.
 
\end{document}